\documentclass[10pt,a4paper]{article}
\usepackage[utf8x]{inputenc}
\usepackage{ucs,color,enumerate}
\usepackage[english]{babel}
\usepackage{amsmath}
\usepackage{amsfonts}
\usepackage{amssymb}
\usepackage[left=3cm,right=3cm,top=3cm,bottom=3cm]{geometry}
\usepackage[round]{natbib}
\usepackage{url}
\usepackage{color,hyperref,enumerate}

\usepackage{graphicx,bbm}
\usepackage{hhline}
\usepackage{subfigure}
\usepackage{bbm}

\title{Semi-Parametric Survival Estimation for pedigrees}

\author{F. Alarcon$^1$, V. Perduca$^1$, G. Nuel$^2$}
\author{Flora Alarcon$^1$, Gregory Nuel$^2$, Violaine Plant\'e-Bordeneuve$^{3,4}$}
\begin{document}

%%%%%%%%%%%%%%%%%%%%%%%%%%%%%%%%%%%%%%%%%%%%%%%%%%%%%
% Dans cette version, on compare les résultats de la plus petite p-value associé au coefficient du snp étudié
% Même pour Plink qui peut prendre en compte l'intéraction
%%%%%%%%%%%%%%%%%%%%%%%%%%%%%%%%%%%%%%%%%%%%%%%%%%%

\maketitle

\hspace{-0.65cm}
$^1$ Mathématiques appliquées Paris 5 (MAP5)
CNRS : UMR8145 -- Université Paris Descartes -- Sorbonne Paris Cité, Paris, France\\
$^2$ Institute of Mathematics, National Center for French Research, Laboratory of Probability, Université Pierre et Marie Curie, Sorbonne Université, France. \\
$^3$ Hôpital Universitaire Henri Mondor, Département de Neurologie
Créteil, France\\
$^4$ Inserm, U955-E10, Créteil, France\\

%%%%%%%%%%%%%%%%%%%%%%%%%%%%%%%%%%%%%%%%%%%%%%%%%%%%%%%%%%%%%%%%%
%%%%%%                                                        Abstract                                      %%%%%%%%%%%%%%%%%%%%%%%%%%%%%%%%%%%%%%%%%%%%%%%%%%%%%%%%%%%%%%%%%%%%%%%%%%%%%%%%%%%%%%%%%%

\begin{abstract}
Mendelian diseases are determined by a single mutation in a given gene. However,
in the case of diseases with late onset, the age at onset is variable; it can even be the case that the onset is not observed in a lifetime. Estimating  the survival function of the mutation carriers and the effect of modifying
factors such as the sex, mutation, origin, etc, is a task of importance, both for management of mutation carriers and for prevention. 
In this work, we present a semi-parametric method based on a proportional to estimate the survival  function using pedigrees ascertained through affected individuals (probands). Not all members of the pedigree need to be genotyped.
The ascertainment bias is corrected by using only the phenotypic information from the relatives of the proband, and not
of the proband himself. The method manage ungenotyped individuals through belief propagation in Bayesian networks and uses an EM algorithm to compute a Kaplan-Meier estimator of the survival function.
The method is illustrated on simulated data and on a samples of families with transthyretin-related hereditary amyloidosis, a rare autosomal dominant disease with highly variable age of onset.
\end{abstract}

%\keywords{Semi-Parametric Survival function, Kaplan-Meier estimator, familial data, Believe Propagation}

\maketitle

%\footnotetext[2]{Please ensure that you use the most up to date
%class file,
%available from the SIM Home Page at\\
%\href{http://www.interscience.wiley.com/jpages/0277-6715}{\texttt{www.interscience.wiley.com/jpages/0277-6715}}}

%%%%%%%%%%%%%%%%%%%%%%%%%%%%%%%%%%%%%%%%%%%%%%%%%%%%%%%%%%%%%%%%%
%%%%%%                                                        Introduction                                   %%%%%%%%%%%%%%%%%%%%%%%%%%%%%%%%%%%%%%%%%%%%%%%%%%%%%%%%%%%%%%%%%%%%%%%%%%%%%%%%%%%%%%

\section{Introduction}
Nowadays, genes involved in the monogenic disease have been identified (e.g. Huntington disease \cite{gusella1983polymorphic, hdcrg1993huntington}). However, in case of monogenic disease with variable age of onset, a precise estimation of the survival function for mutation carrier individual is necessary as well as identification of potential factors  that modulate this age. Indeed, this estimates allows to provide individual risk assessment, to understand the underlying mechanisms of the disease and to establish prevention strategies. Sometimes, the age at onset can be so late that a significant proportion of mutation carriers do not declare the disease in the lifetime. We called this phenomenon the incomplete penetrance. It should be noted that in this literature, the age-specific cumulative distribution function (CDF), named also penetrance function, is preferentially evoked. In this paper, we will use the classical survival function (which is simply the complementary of the CDF).

Because of the low carriage frequency and the hight cost of genetic test, random sampling is not a praticable approach to obtain a sample of sufficient size to draw reliable conclusions. Data are usually obtained from families ascertained through affected individuals. Indeed, as all affected individual necessary carry the mutation,  the families ascertained in this way are informative to estimate the survival function. The drawback of this procedure is that the survival function can be significantly underestimated if the ascertainment process is not taking into account \cite{carayol2002hnp}.  Therefore, an adjustment for the ascertainment bias is required.

The ascertainment correction problem is a very challenging problem. Vieland and Hodge explain in their articles \cite{vieland1996problem, vieland1995inherent} that ``\emph{without knowledge of the true underlying pedigree structure (including who are the unobserved members of pedigree) it is not possible to write down a correct likelihood and the ascertainment correction problem becomes intractable}''. However, different adjustments for ascertainment have already been suggested to provide valid risk estimate of a genetic disease \cite{carayol2004epf, lebihan1995ame, alarcon2009pel}. 

In monogenic disease, as all affected carry the mutation, an ascertainment through affected individuals is sufficient to have mutation carriers.
When pedigree are ascertained through at least one affected individual, it is possible to correct ascertainment bias by modeling analytically the ascertainment correction \cite{kraft2000bae, plantebordeneuve2003gst}. However, this prospective correction require additional parameters as $\pi$, the probability for an affected to be ascertained, which have to be estimated and make the strong assumption that all affected have the same chance to be ascertained.

An other more intuitive method, the PEL, have been proposed that corrects for ascertainment by simply removing the phenotypic information of the individual (called \textit{proband}) who allowed his family to be selected \cite{alarcon2009pel} (i.e. proband's phenotype exclusion). A similar method had been proposed by Weinberg \cite{weinberg1912methoden, weinberg1928mathematische} to correct for the ascertainment bias in the estimation of segregation ration (see also \cite{fisher1934effect}). Moreover, it has been shown (in \cite{alarcon2009pel}) for various genetic models and selection schemes that PEL corrects better than the prospective method.

The PEL is a parametric method in which the age at onset is modeled by a Weibull distribution. Although this model is widely used in survival analysis because of its capacity to adjust to observed data, it can fail to fit properly the survival function in some cases.
The advantage of a non-parametric estimation of the survival function has been shown in \cite{alarcon2009ideal}, as well as the ability of the proband's phenotype exclusion to correct for the ascertainment bias in this context. However, the method proposed in \cite{alarcon2009ideal} assume tha all genotypes are observed, which is a strong hypothesis that prevent any application on real data set.

In this article, we introduce a semi-parametric method based on a proportional hazard (Cox model) to estimate survival function from familial data. The presence of ungenotyped individual in the families are managed through belief propagation in Bayesian networks which allows to estimate, for all unaffected individual,  his probability to be a carrier. This probability is then taken into account in the survival function estimation through weights.

 The method uses an EM algorithm to compute a Kaplan-Meier estimator of the survival function and correct for the ascertainment bias by excluding the proband's phenotype, like in the PEL. An other advantage of this non parametric method is its ability to accommodate covariates (as sex, mutation, etc.) thanks to the Cox model.
The EM algorithm can be summarized as follow : 1) Survival function are estimated with arbitrary weight standing for the individual probability to carry the mutation in the M-step. 2) Then the weight is assessed according to the estimated survival function in the E-step.

Section 2 presents first the estimation model and then it describes the believe propagation in this contexte  as well as the E-M algorithm. Section 3 presents  the results obtained on simulated data sets and Section 4 illustrates the method on transthyretin-related hereditary amyloidosis families from different origin (French, Portuguese, and Swedish). For French families, two different mutations are compared through a log-rank test whereas our non-parametric estimation is compared with a Weibull estimation in Portuguese dataset. Finally, methodology and results are discussed in the Section 5. 

 %%%%%%%%%%%%%%%%%%%%%%%%%%%%%%%%%%%%%%%%%%%%%%%%%%%%%%%%%%%%%%%%
%                                           Semi-parametric estimation of the survival function                                                                               %%%%%%%%%%%%%%%%%%%%%%%%%%%%%%%%%%%%%%%%%%%%%%%%%%%%%%%%%%%%%%%%

\section{ Semi-parametric estimation of the survival function }

%%%%%%%%%%%%%%%%%%%%%%%%%%%%%%%%%
%%%%%%%%%%%%%%%%%%%%%%%%%%%%%%%%%
\subsection{The model}

Let's consider $n$ individuals indexed by $i=1, \dots, n$. For an individual $i$, we denote by $(T_i,\delta_i)$ the vector defined for $T_i \leq 0$ and $\delta_i \in \{0,1\}$ as follows : 

$$
T_i=\left\{
\begin{array}{ll}
\text{age at diagnosis} & \text{if $\delta_i=1$} \\
\text{age at last follow-up} & \text{if $\delta_i=0$} \\
\end{array}
\right.
$$

We denote by $X_i \in \{00,01,10,11\}$, the genotype of individual $i$ where the first number represents the number of disease allele ($\in \{0,1\}$)  transmitted by the father and the second one represents the number of disease allele ($\in \{0,1\}$) transmitted by the mother. So $X_i=01$ means that the individual $i$ carry the mutation, that he is heterozygous and that his mutation have been transmitted to him by his mother. Note that this variable is often unobserved because individual are rarely genotyped, and that distinguishing betwee, ``01'' and ``10'' usually requires to take into account the whole pedigree structure.

Finally, we consider the vectors of dimension $n$ of the sample : $\boldsymbol T = (T_1,\ldots,T_n)$, $\boldsymbol \delta = (\delta_1,\ldots,\delta_n)$, $\boldsymbol X = (X_1,\ldots,X_n)$, and we assume the following model:
$$
\mathbb{P}(\boldsymbol T, \boldsymbol \delta , \boldsymbol X)=
\underbrace{\prod_{i=1}^{n} \mathbb{P}(T_i, \delta_i |  X_i )}_{\text{survival part}}
\times 
\underbrace{\prod _{i=1}^{n} \mathbb{P}\left (X_i |  X_{\text{father}_i} , X_{\text{mother}_i} \right)}_{\text{genetic part}}
$$
where $\text{father}_i$ and $\text{mother}_i$ indicate the father and mother of individual $i$ (empty information for the founders). 
We will now detail both the survival and the genetic part of this model.

%%%%%%%%%%%%%%%%%%%%%%%%%%%%%%%%%%%%%

\subsubsection{Survival part.}

As we know that all affected individuals are necessarily carriers, we assume that non carrier \emph{cannot} be affected:
$$
 \mathbb{P}(T_i, \delta_i |  X_i = 00) = 1
$$
and we consider a dominant model with incomplete penetrance and proportional hazards (PH):
$$
\log  \mathbb{P}(T_i, \delta_i |  X_i \neq 00) =-\Lambda(T_i) e^{\boldsymbol Z_i \boldsymbol \beta} + \delta_i \left( \lambda(T_i) + \boldsymbol Z_i \boldsymbol \beta \right)
$$
where $\lambda(t)$ is the baseline hazard, $\Lambda(t)= \int_{0}^{t} \lambda(u)du$ is the baseline cumulative hazard, $\boldsymbol{Z}_i$ are the covariates of individual $i$, and $\boldsymbol \beta$ is a regression coefficient.

In this paper, we will be led to consider two forms for  $\lambda$ : 

\begin{enumerate}[i)]
\item Weibull: $\lambda(t)$ is the density of a Weibull distribution; a classical parametric choice in the context of survival analysis. Other popular choices include the exponential or log-gamma distributions.
\item Nelson–Aalen: a non-parametric piecewise constant baseline survival $S(t)=\exp(-\Lambda(t))$; the classical non-parametric choice in survival analysis (NB: in the particular case where there is no PH model, this estimator is due to Kaplan-Meier and is therefore often improperly referred under this name even in the PH case).
\end{enumerate}

%%%%%%%%%%%%%%%%%%%%%%%%%%%%%%%%%%%%%

\subsubsection{Genetic part.}

We assume a classical genetic model : Hardy-Weinberg equilibrium is assumed in pedigree founders and the disease allele frequency $q$ is assumed to be known for the founders. Moreover, Mendelian transmission of the alleles from parents to offspring is assumed. Since our $n$ individuals might belong to completely independent families, it is clear that the genetic likelihood can be computed separately on this independent families, however, the notations are still valid, dramatically simpler by ignoring the family level.

Another important point is the fact that the true genotype $X_i$ is at best partially observed. Indeed, a positive mutation search or an affected individual, only indicates that $X_i \neq 00$ is impossible. On the other hand, a negative mutation search indicates that $X_i=00$ (assuming a 100\% sensitivity of the mutation search procedure). More complex model allowing for genotyping errors or even pedigree errors (wrong filiation for example) can be incorporated like in \cite{thomas2005gmcheck}. In the present work, we decided to use the most basic (but reasonable) model.

%%%%%%%%%%%%%%%%%%%%%%%%%%%%%%%%%
%%%%%%%%%%%%%%%%%%%%%%%%%%%%%%%%%
\subsection{The EM algorithm}

Since $\boldsymbol X$ is only partially observed, we consider this variable as latent and use a classical Expectation-Maximization algorithm in order to maximize the model log-likelihood in parameter $\boldsymbol \theta=(\lambda,\boldsymbol \beta)$ (allele frequency $q$ is assumed to be known). For this purpose, we first need to incorporate the auxiliary $Q$ function:
$$
Q\left( \boldsymbol \theta | \boldsymbol \theta_\text{old} \right) \stackrel{\text{def}}{=}
\int \mathbb{P}\left(\boldsymbol X | \text{ev};\boldsymbol\theta_\text{old}\right) \log \mathbb{P}(\boldsymbol{T},\boldsymbol{\delta},\boldsymbol{X}; \theta) d \boldsymbol X
= \sum_{i=1}^{n} w_i\log \mathbb{P}(T_i,\delta_i | X_i \neq 00)
$$
where $\text{ev}$ denote the evidence (that means $\boldsymbol{T}$, $\boldsymbol{\delta}$, and any mutation search information), and where the weights $w_i$ are defined as:
$$
w_i \stackrel{\text{def}}{=}\mathbb{P}\left( X_i \neq 00  | \text{ev};\boldsymbol\theta_\text{old}\right)
$$

%%%%%%%%%%%%%%%%%%%%%%%%%%%%%%%%%%%%
\subsubsection{E-step}

The auxiliary function $Q$ is computed at this step. In our case, the marginal weights $w_i$ are all we need to compute. Due to the complex dependency structure of the genotype in our pedigree, this is however a challenging task. In the particular case where the pedigree is a simple tree, the classical Elston-Stewart algorithm can be used \cite{elston1971general}. When loops are present (consanguinity, mating loops, twins), Elston-Stewart must be combined with loop breaking approaches at the cost of an exponentional complexity with the number of loops \cite{lange1975extensions}. Instead of Elston-Stewart we consider here the belief propagation algorithm (also called sum-product algorithm) in Bayesian network which can be see as a generalization\footnote{even if belief propagation was developed independently by the probabilist graphical model community.} of Elston-Stewart to arbitrary pedigrees. See Section 2.3 for more details on belief propagation.

The only information we need to provide for this step is called the \emph{evidence} $\text{ev}$ and is defined as:
$$
\text{ev}_i(x) = 
1_{\{\text{$X_i=x$ compatible}\}} \times \left\{ 
\begin{array}{ll}
1 & \text{if $\delta_i=1$} \\
\mathbb{P}(T_i,\delta_i=0 | X_i = x ) & \text{if $\delta_i=0$} 
\end{array}
\right.
$$
(ex: $x\neq 00$ is incompatible with a negative mutation search on an affected individual). Note that the evidence $1$ for affected individual can be used because the $T_i$ is non-informative for the distribution of $X_i$ in this particular case. This is also better for our non-parametric estimation which cannot provide hazard estimate without smoothing (ex: kernel smoothing) but only survival estimates.

\subsubsection{M-step}

The auxiliary function $Q$ is maximized at this step. As seen above, our auxiliary function can be simply seen as the classical log-likelihood of a survival model where each individual observation receive the weight $w_i$.  This is hence a very classical problem which can easily be handled via classical statistical optimization procedure. Under the programming software R \cite{team2014ra} and the \texttt{survival} package \cite{therneau2013package, therneau2000modeling} one can use:
\begin{itemize}
\item proportional hazard model: $\texttt{coxph}$ function with optional $\texttt{weights}$ $w_i$. This procedure can easily handle any form of covariates, including stratification. Significance of regression coefficients can be reported.
%, and even frailty (ie: random effect) but is known to be non-robust for that last task\flo{ On laisse cette phrase???? ou on met plutôt un truc du genre "frailty have not be included in our model and will be discussed in the Section 5"}. 

\item parametric survival estimates: $\texttt{survreg}$ with Weibull distribution (or other distributions), directly applied to a $\texttt{coxph}$ object where we specify the covariate nature of the baseline.
\item non-parametric survival estimates: $\texttt{survfit}$ directly applied to a $\texttt{coxph}$ object where we specify the covariate nature of the baseline. Time specific confidence interval are provided.
\end{itemize}

\subsubsection{Practical implementation}

Initialization is performed by affecting random weights $w_i$ (ex: drawn from a uniform distribution on $[0,1]$). EM iterations are stopped when we observe convergence on test survival estimates (ex: baseline survival at age $20,40,60,80$). 

%%%%%%%%%%%%%%%%%%%%%%%%%%%%%%%%%%%%%%%%
%%%%%%%%%%%%%%%%%%%%%%%%%%%%%%%%%%%%%%%%

\subsection{Believe propagation in pedigrees}
%
%The sum in the likelihood over all possible genotypic configuration is tackled by believe propagation....
%
%This computation need to know an additional parameter $q$ that is the frequency of the disease allele in the population. As our families are ascertained though affected individuals, we can not use our sample to estimate $q$. Hence, we do not consider $q$ as a parameter of $L$ ; we fix its value according to expert knowledge or from estimations of the disease prevalence. 
%

\begin{figure}
\begin{center}
\begin{tabular}{cc}
\includegraphics[width=0.32\textwidth]{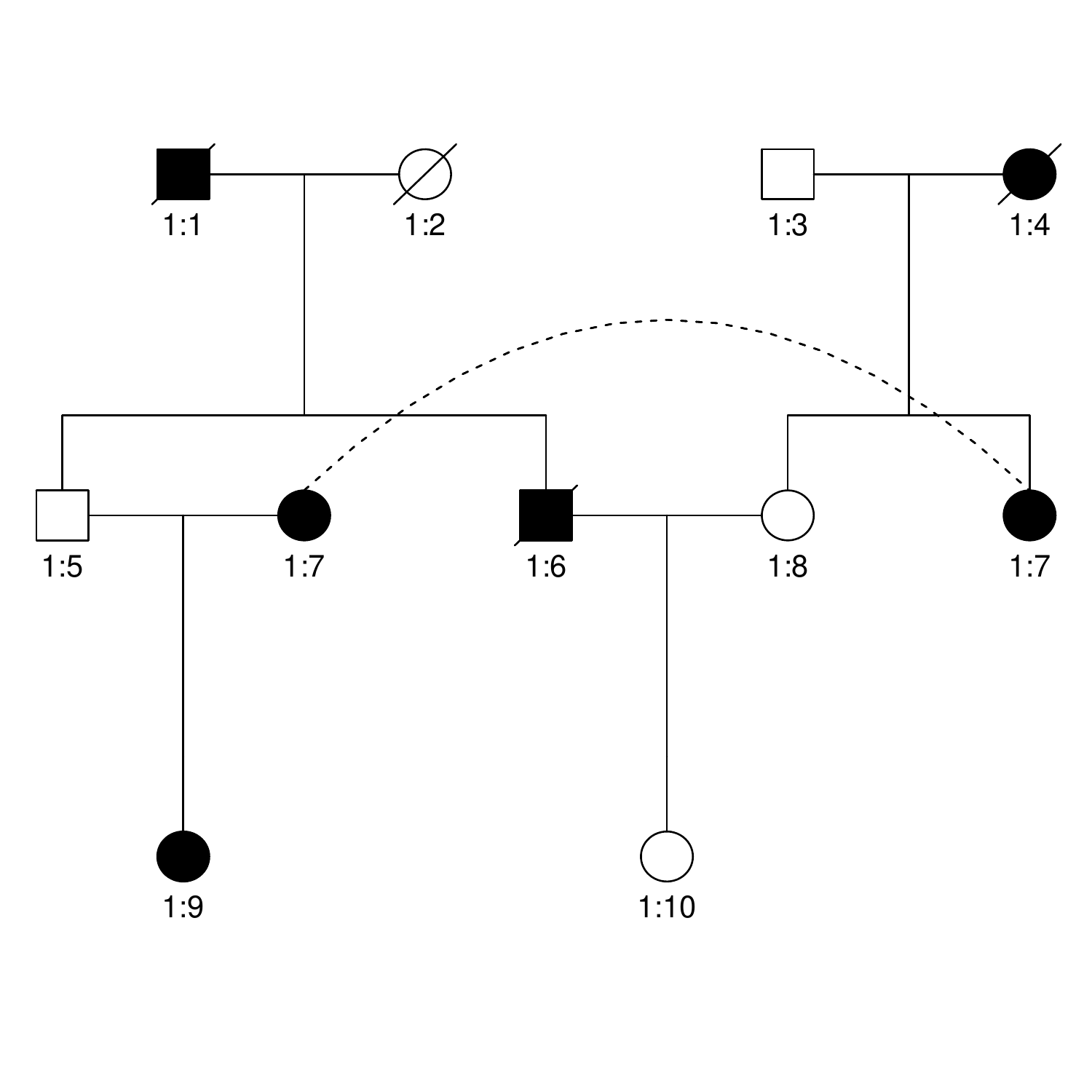}&
\includegraphics[width=0.68\textwidth]{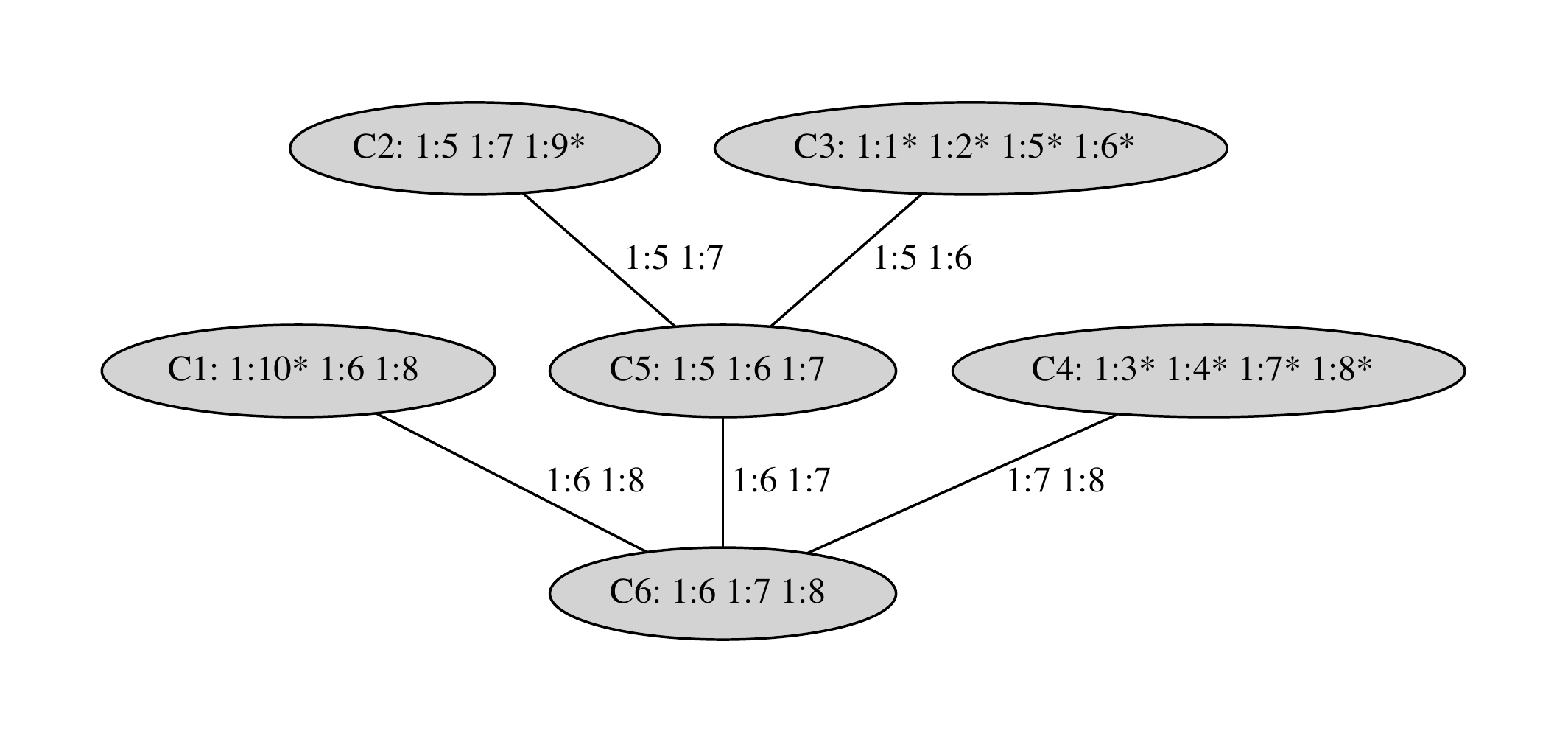}\\
(a) & (b)
\end{tabular}
\end{center}
\caption{The ``mating loop'' example. Panel a: the pedigree (the dashed line indicates that individual ``1:7'' is represented twice in the pedigree but is in fact a single individual); Panel b: a junction tree corresponding to the problem (the `*' indicate the locations where individual evidences are injected into the junction tree).}\label{fig1}
\end{figure}
%%%%%%%%%%%%%%%%%%%%%%%%%%%%%%%%%%%%%%%%%%%%%%%%%%%%%%%
%%%%%%%%%%%%%%%%%%%%%%%%%%%%%%%%%%%%%%%%%%%%%%%%%%%%%%%
Belief propagation (BP) in pedigree is a very general method which can deal efficiently with very complex pedigree structure (ex: $2000$ individuals with $50$ loops). Unlike Elston-Stewart algorithm, BP does not use loop breaking approaches to deal with loop pedigrees. Instead, BP use an auxiliary tree called the junction tree (JT) which basically is a clique decomposition of the moral graph corresponding to the pedigree problem. JT and BP are well known is the graph theory (ex: JT can be used to solve a graph coloring problem) and in the mathematical field of probabilistic graphical models (Bayesian network, hidden Markov model, decision trees, Markov networks, etc.).

In Figure~\ref{fig1}a we represent a simple example pedigree with a mating loop. This is typically a pedigree which would require to perform loop breaking (for example on $1:7$) in order to be solved by Elston-Stewart. Here we build instead the JT of Figure~\ref{fig1}b in which the \emph{evidence} (see above) is injected prior to the BP. Then BP consists in computing and propagating recursively so-called \emph{messages} from the leaves to the root. Here we use evidence of $1:10$ to compute $M_{1,6}$ from $C_1$ to $C_6$, then evidence of $1:9$ for $M_{2,5}$, evidence of  $1:1$, $1:2$, $1:5$ and $1:6$ for $M_{3,5}$, then $M_{2,5}$ and $M_{3,5}$ for $M_{5,6}$, then evidence $1:3$, $1:4$, $1:7$ and $1:8$ for $M_{4,6}$, and finally $M_{1,6}$, $M_{5,6}$, and $M_{4,6}$ at the root. After this \emph{inward} propagation, evidence can be recursively propagated back to the leaves (\emph{outward} propagation) in order to obtain marginal posterior distribution of the variables. 

\begin{table}
$$
\begin{array}{ccccc}
  \hline
i  & x=00 & x=10 & x=01 & x=11 \\
  \hline
1:1 & 0.000 & 0.494 & 0.494 & 0.012 \\ 
1:2 & 0.965 & 0.017 & 0.017 & 0.000 \\ 
 1:3 & 0.965 & 0.017 & 0.017 & 0.000 \\ 
 1:4 & 0.000 & 0.495 & 0.495 & 0.010 \\ 
1:5 & 0.389 & 0.591 & 0.009 & 0.012 \\ 
1:6 & 0.000 & 0.977 & 0.012 & 0.012 \\ 
1:7 & 0.000 & 0.010 & 0.975 & 0.016 \\ 
1:8 & 0.486 & 0.009 & 0.496 & 0.009 \\ 
1:9 & 0.000 & 0.203 & 0.590 & 0.207 \\ 
1:10 & 0.365 & 0.374 & 0.129 & 0.132 \\ 
   \hline
\end{array}
$$
\caption{Posterior distribution $\mathbb{P}(X_i=x | \text{ev})$ computed though BP for the ``mating loop'' example.}\label{tab1}
\end{table}

Let us see what give BP on our example assuming that allele frequency is $q=1\%$ and that all affected are carrier (no other information is provided). The posterior marginal distribution for all individuals in the pedigree is given in Table~\ref{tab1}. Without surprise, we observe that all affected individuals ($i=1,4,6,7,9$) cannot have the non-carrier genotype $00$.  If we look to individual $1:4$, she has genotypes $10$ or $01$ with equal probability $0.495$ and hence, she can be an homozygous carrier with probability $0.01$, the allele frequency, which is consistent. Now, individual $1:7$ is also a carrier, but the fact that her mother is indeed a carrier makes much more likely that her genotype is $01$, and this is clearly accounted by the BP.

%%%%%%%%%%%%%%%%%%%%%%%%%%%%%%%%%%%%%%%%%%%%%%%%%%%%%%%%%%%%%%%%%
%                                              Analysis of simulated datasets                                                                      					    %%%%%%%%%%%%%%%%%%%%%%%%%%%%%%%%%%%%%%%%%%%%%%%%%%%%%%%%%%%%%%%%%

\section{Analysis of simulated datasets}

%%%%%%%%%%%%%%%%%%%%%%%%%%%%%%
%%%%%%%%%%%%%%%%%%%%%%%%%%%%%%
\subsection{Simulation of pedigrees}
We simulated families with realistic size and structure from 35 french families with transthyretion-related hereditary amyloidosis (see the next section on analysis of real data). We duplicated families $k$ times (here we fixed $k=3$) in order to have a larger sample of 105 families. Ages, sex and the proband individual in each families is given by the real dataset.
Genotypes are assigned respecting Mendelian transmission, with a disease allele frequency $q=0.004$ in our simulated dataset. The age at event is simulated according to a piecewise constant hazard rate function, $\lambda(t)$, given as follows and an uniform censoring variable is added. Finally, only families with at least one affected individual is retained in the dataset, representing the ascertainment.
$$
\lambda(t)=\left\{
\begin{array}{ll}
0 & \text{ if } t \in [0, 20]\\
0.02 & \text{ if } t \in [20, 40]\\
0.10 & \text{ if } t \in [40, 60]\\
0.05 & \text{ if } t > 60 
\end{array}
\right.
$$

The last step of the simulation is to set all genotypes to ``unknown'', so that all simulated data are analysed without knowledge of the genotypes.

%%%%%%%%%%%%%%%%%%%%%%%%%%%%%%
%%%%%%%%%%%%%%%%%%%%%%%%%%%%%%
\subsection{Method assessment on a simple simulated dataset}

\begin{figure}[htp]
\centering
\includegraphics[height=0.6\textwidth]{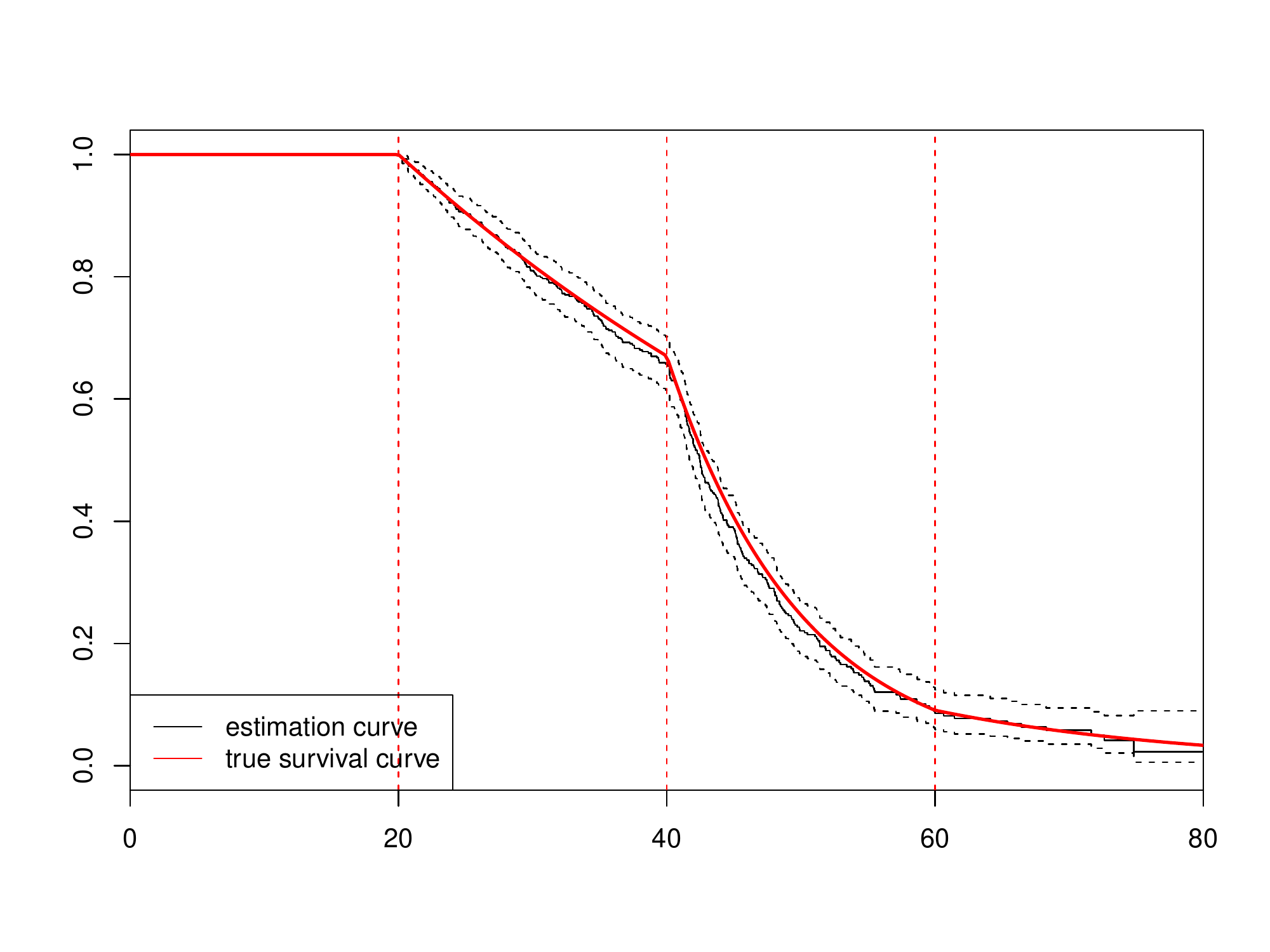}
\caption{Estimation of the baseline survival function $S_0(t)$ with confidence intervals for a simulated dataset}\label{ssd}
\end{figure}

We first assessed the method on a simple simulated dataset without regardless additional covariates in the model. Figure \ref{ssd} shows that our method succeed in estimating the true survival curve (red curve) even if all genotypes are considered as missing. Furthermore, the sample size leads to smaller confidence intervals. These first result allow us to validate the ascertainment bias correction since families who have disease mutation but no affected individual are not ascertained.

%%%%%%%%%%%%%%%%%%%%%%%%%%%%%%
%%%%%%%%%%%%%%%%%%%%%%%%%%%%%%
\subsection{ Stratification or proportional hazard to take account for covariate}
\begin{figure}[htp]
\centering
\includegraphics[height=0.6\textwidth]{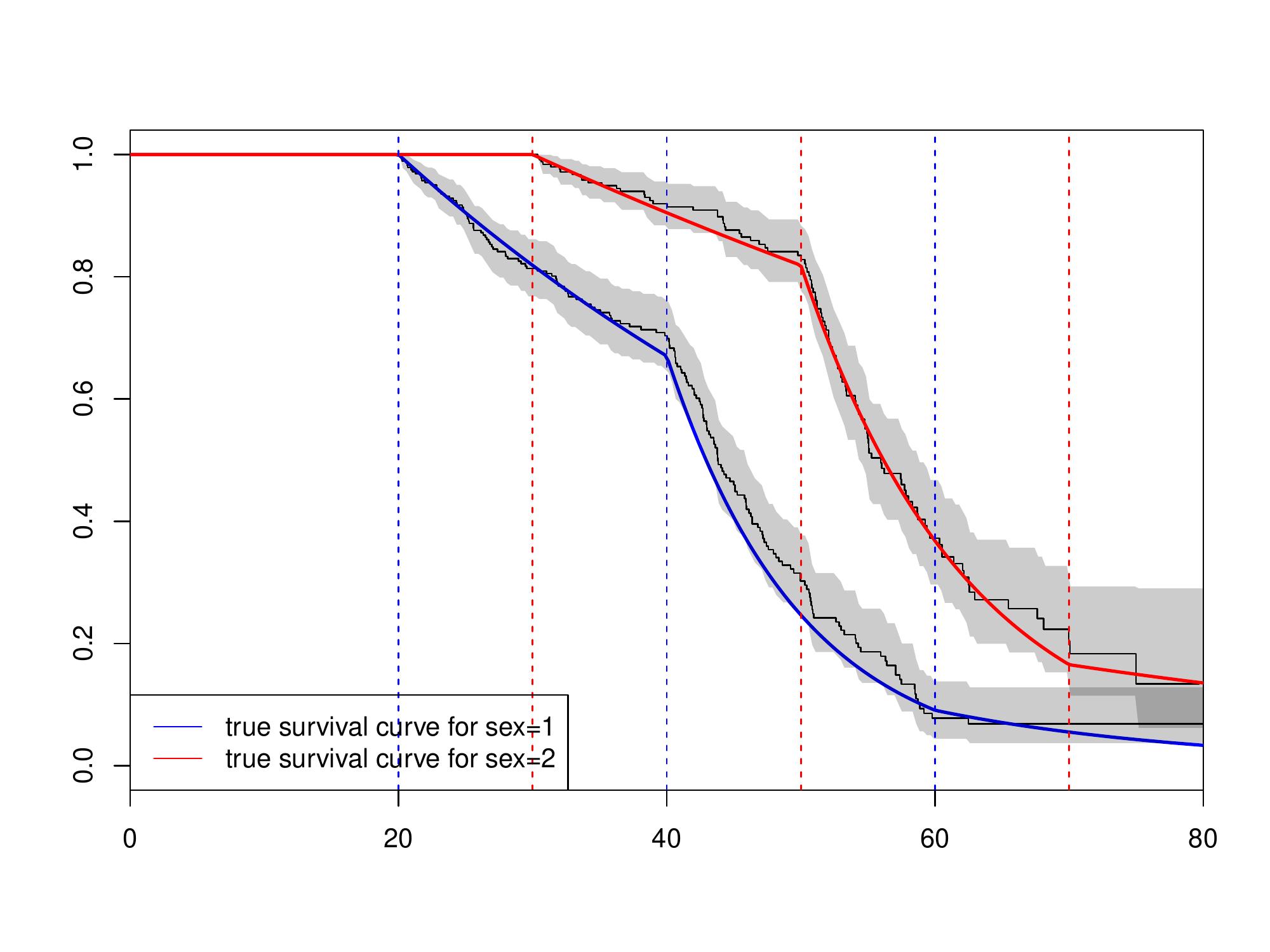}
\caption{Estimation of Survival curve with stratified sex effect}\label{stratsd}
\end{figure}

\begin{figure}[htp]
\centering
\includegraphics[height=0.6\textwidth]{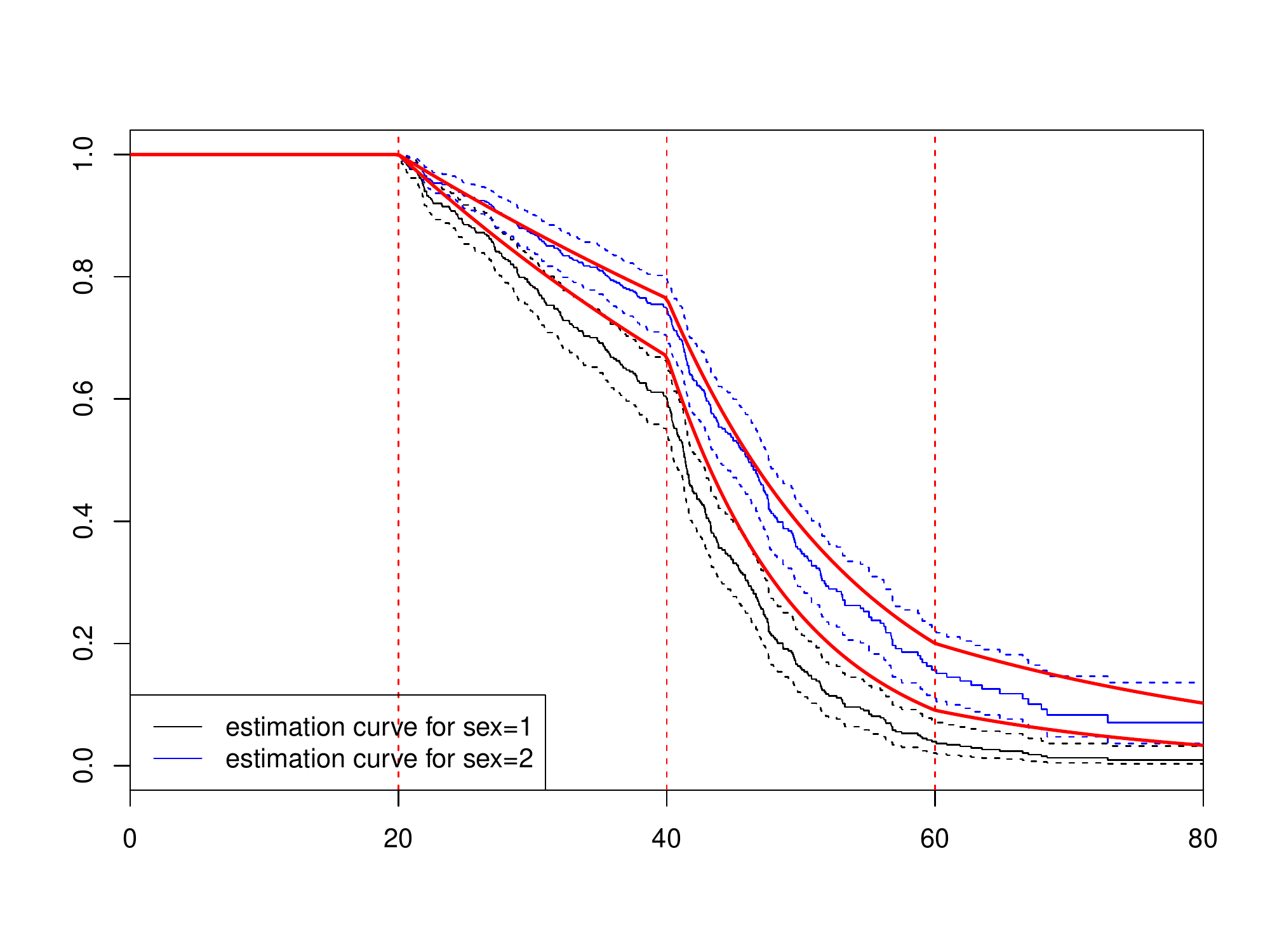}
\caption{Estimation of Survival curve with a proportional sex effect}\label{propsd}
\end{figure}

When covariates are available, we have the choice between stratify on these covariates or take account on the covariate in the proportional hazard model. Figure \ref{stratsd} shows estimation of survival curve (black lines) stratified for males (sex=1) and females (sex=2) in a simulation framework where the age at diagnosis has been simulated according to different piecewise constant hazard rate depending on the sex of the individual. 95\% confidence intervals are provided through polygons. Figure \ref{propsd} shows estimation of the survival curve (with 95\% confidence interval in dotted lines) for males and females when a proportional protector effect of the female sex are added in simulations through a cox model. The $\beta$ parametric parameter of the model have set to $-0.4$. Thus, the women's survival curve is higher than men's. Here the $\beta$ parameter was estimated by $\hat{\beta} = -0.56$

%%%%%%%%%%%%%%%%%%%%%%%%%%%%%%%%%%%%%%%%%%%%%%%%%%%%%%%%%%%%%%%%%
%                                              Analysis of real datas                                                                             					    %%%%%%%%%%%%%%%%%%%%%%%%%%%%%%%%%%%%%%%%%%%%%%%%%%%%%%%%%%%%%%%%%

\newpage
\section{Analysis of real data}
We illustrated the method on transthyretin-related hereditary amyloidosis, an autosomal dominant disease, caused by a mutation of the TTR gene, Val30Met (MET30) substitution being the most frequent mutation \cite{plante2011familial}. The age at onset ranges from early twenties to late seventies. Although distributed worldwide, the disease is often clustered in limited areas like in Portugal, Japan and Sweden with different genotypic and phenotypic variation. In France, we are dealing with two populations, i.e. of Portuguese and of French origins. While many pathogenic TTR variants have been detected among French population, only one variant, the MET30, was detected in the Portuguese population.

We analyse three data set constituted of 49 families of French descent, 33 families of Portuguese descent and 78 families of Swedish descent, ascertained through affected individuals (see Table \ref{fam-descr}).
Data are analyzed excluding the proband to avoid ascertainment biais (as done in simulations) and the deleterious allele frequency was arbitrarily set to $q=0.004$ and the \textit{de novo} mutation was set to 0.

\begin{table}[htp]\centering
\begin{tabular}{lrrr}
           & French  & Portuguese & Swedish \\
%\midrule
\hline
All        & 1238 & 1191 & 1361 \\
Affected   &  87 & 178 & 151\\
%\bottomrule
\hline
\end{tabular}
\medskip
\caption{Description of French, Portuguese and Swedish families}\label{fam-descr}
\end{table}

\subsection{A french dataset}
Among the 30 different substitutions of the TTR observed in families of French descent, MET30 and Ser77Tyr (TYR77) are the most frequent accounting for about 50\% of the kindreds. Age at first symptoms is significantly much older than in families of Portuguese descent but appear similar in both variants in the French families.

\begin{figure}[htp]
\centering
\includegraphics[height=0.6\textwidth]{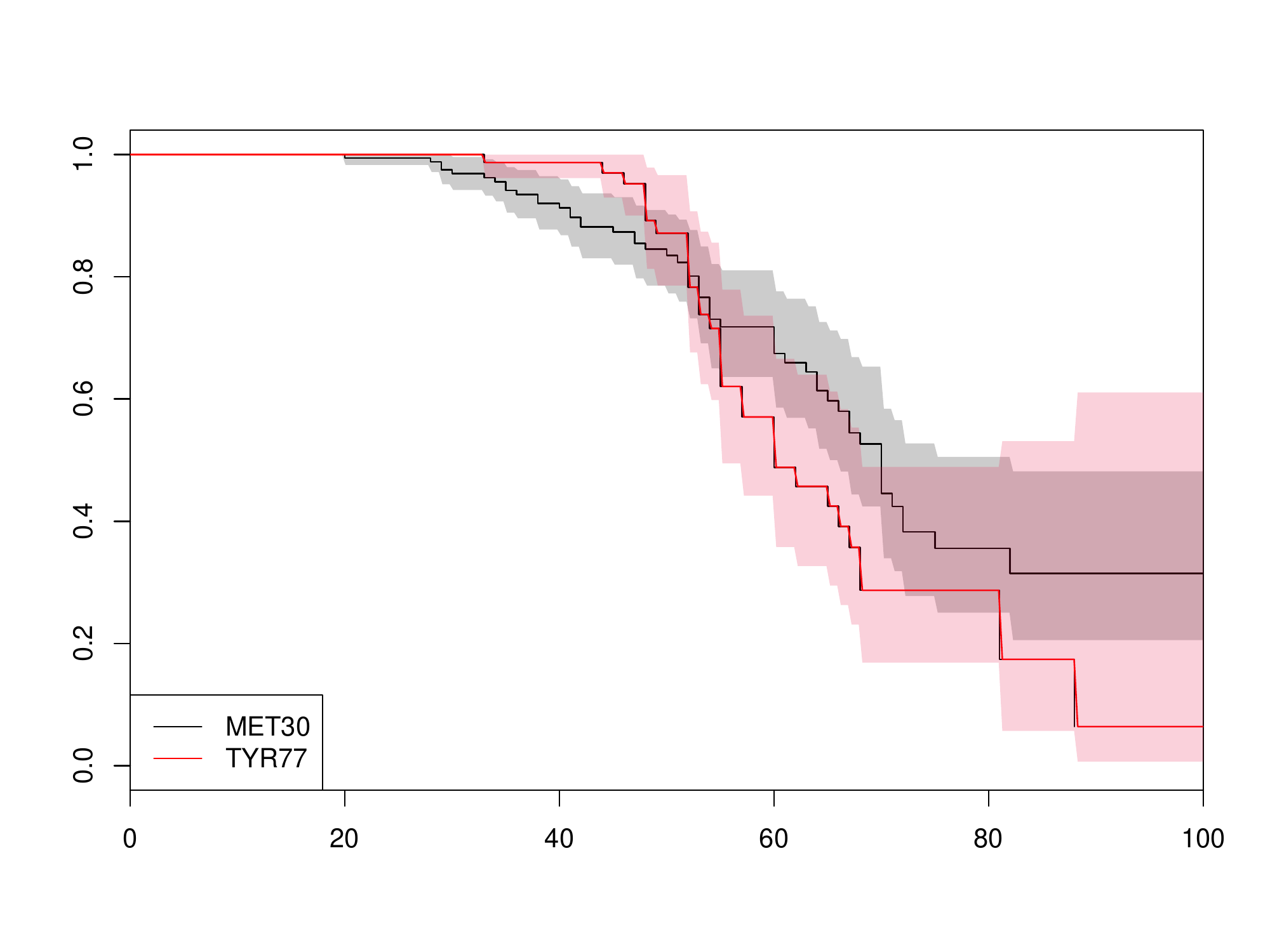}
\caption{Estimation of Survival curve stratified on mutation in the French dataset}\label{frenchmutmettyr}
\end{figure}

We analyse a French dataset affected by transthyretin-related hareditary amyloidosis. The sample set consist in 35 families with a mutation MET30 and 15 families with a mutation TYR77. Figure \ref{frenchmutmettyr} shows the survival curves estimated stratified on the type of mutation effect.
A log-rank test was performed with the R function \textit{surdiff} in order to compare the two mutations. Thus, a significant difference is tested between survival curve for MET30 mutation (black curve) and TYR77 mutation (red curve) with a p-value estimated to $0.002$. 95\% confidence intervals are given through colored regions.

%%%%%%%%%%%%%%%%%%%%%%%%%%%%%%
%%%%%%%%%%%%%%%%%%%%%%%%%%%%%%
\subsection{A Portuguese dataset}

\begin{figure}[htp]
\centering
\includegraphics[height=0.6\textwidth]{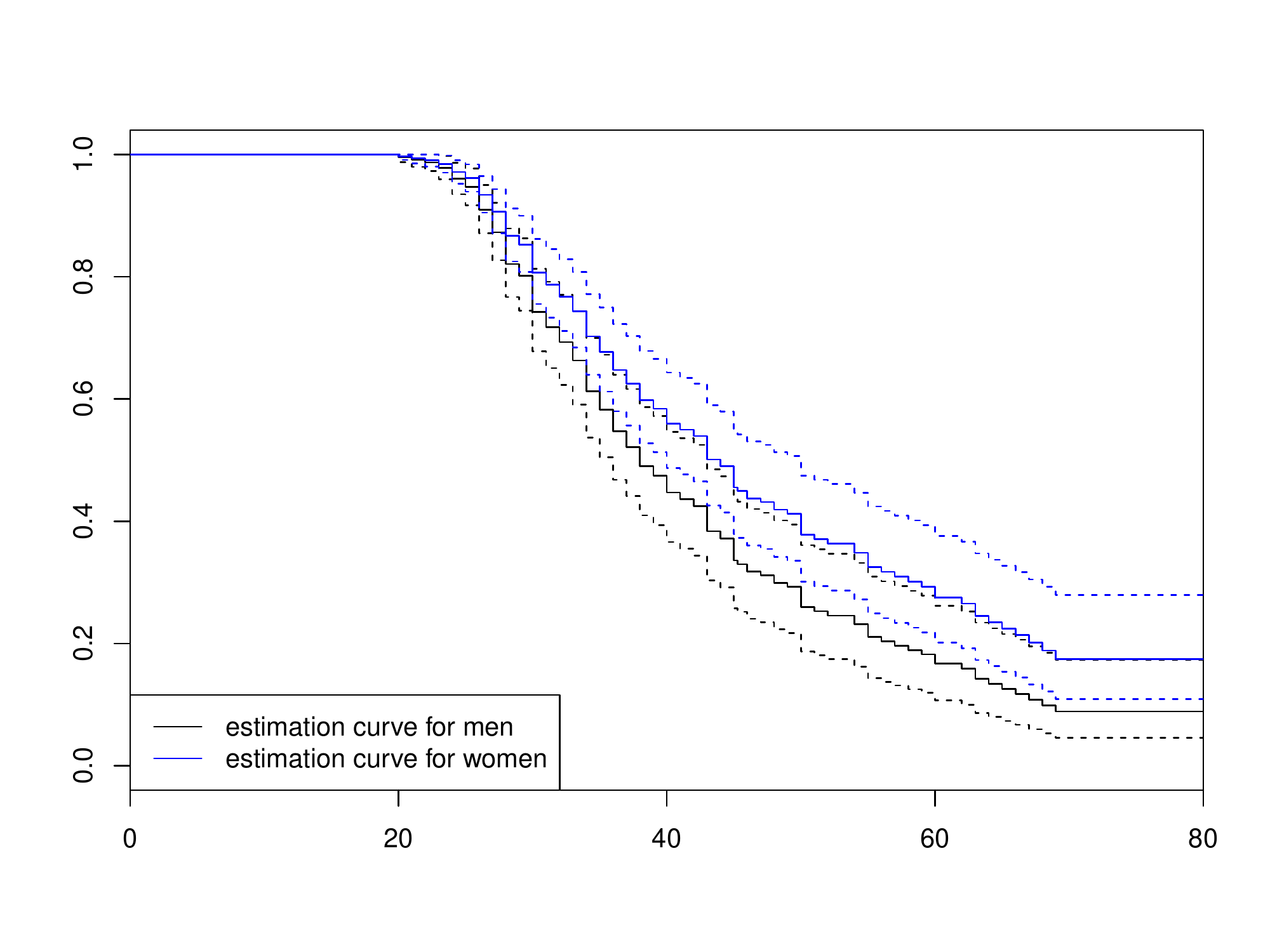}
\caption{Estimation of the survival function with sa proportional sex effect on Portuguese dataset}\label{portsex}
\end{figure}

\begin{figure}[htp]
\centering
\includegraphics[height=0.6\textwidth]{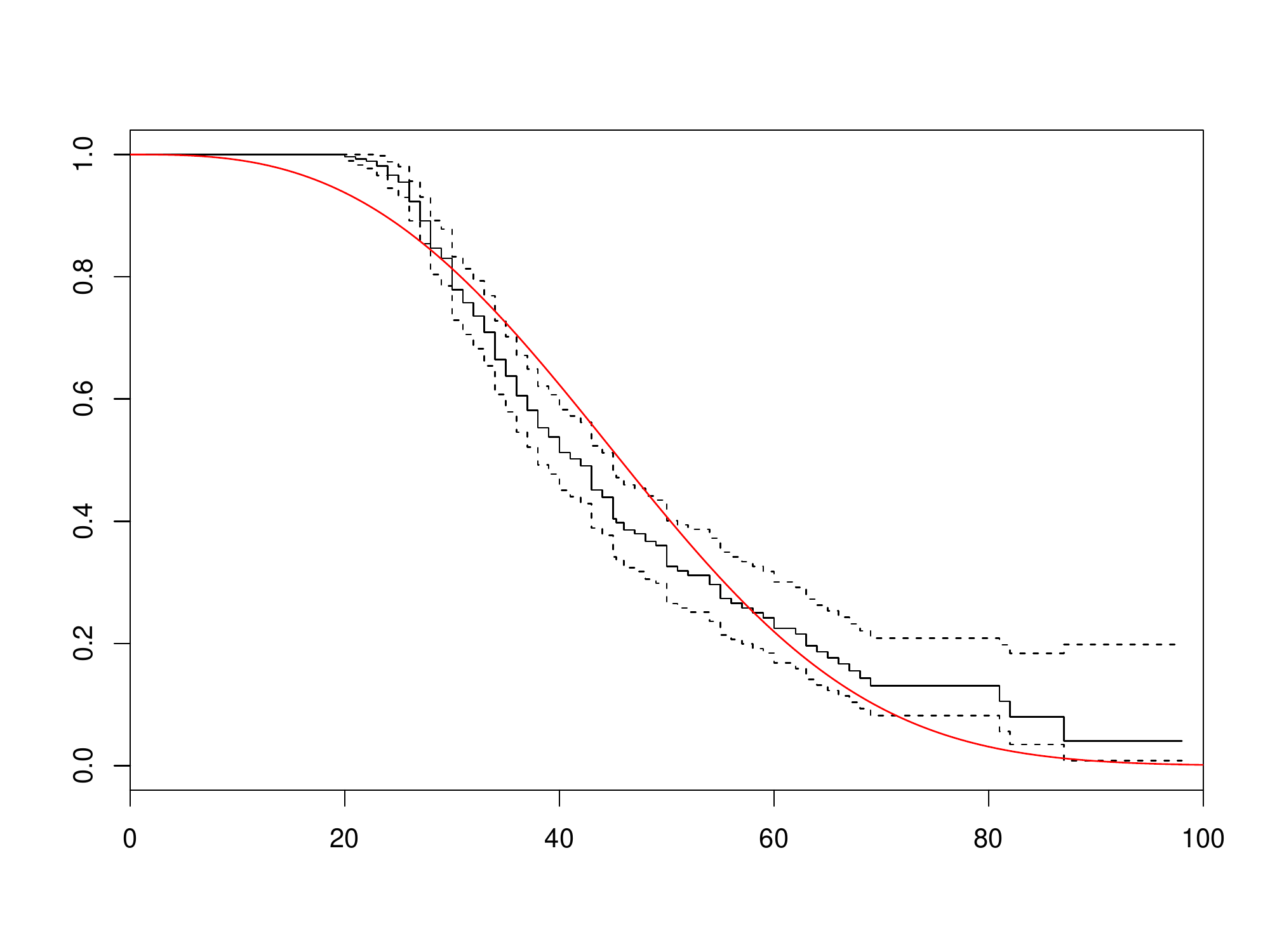}
\caption{Comparison between a Weibull parametric method to estime the Survival curve in the Portuguese dataset}\label{portWeibull}
\end{figure}

In this section, we analyse a data set constituted of 33 families of Portuguese descent, first described in \cite{alarcon2009pel}. 
Figure \ref{portsex} shows the survival curves estimated with a proportional sex effect, given with 95\% confidence intervals. The proportional parameter $\beta$ is estimated to $\hat{\beta} = -0.327$ with a p-value $p = 0.034$. We can note that the survival is lower in Portuguese data set than in French data set. This results have already been shown in \cite{plantebordeneuve2003gst}.
Figure \ref{portWeibull} shows the comparison between our method and a Weibull parametric estimation assessed through a E-M algorithm with the R function \textit{Survreg}. We observe that the Weibull estimation does not fit the non-parametric curve.

\subsection{A Swedish dataset}
\begin{figure}[htp]
\centering
\includegraphics[height=0.6\textwidth]{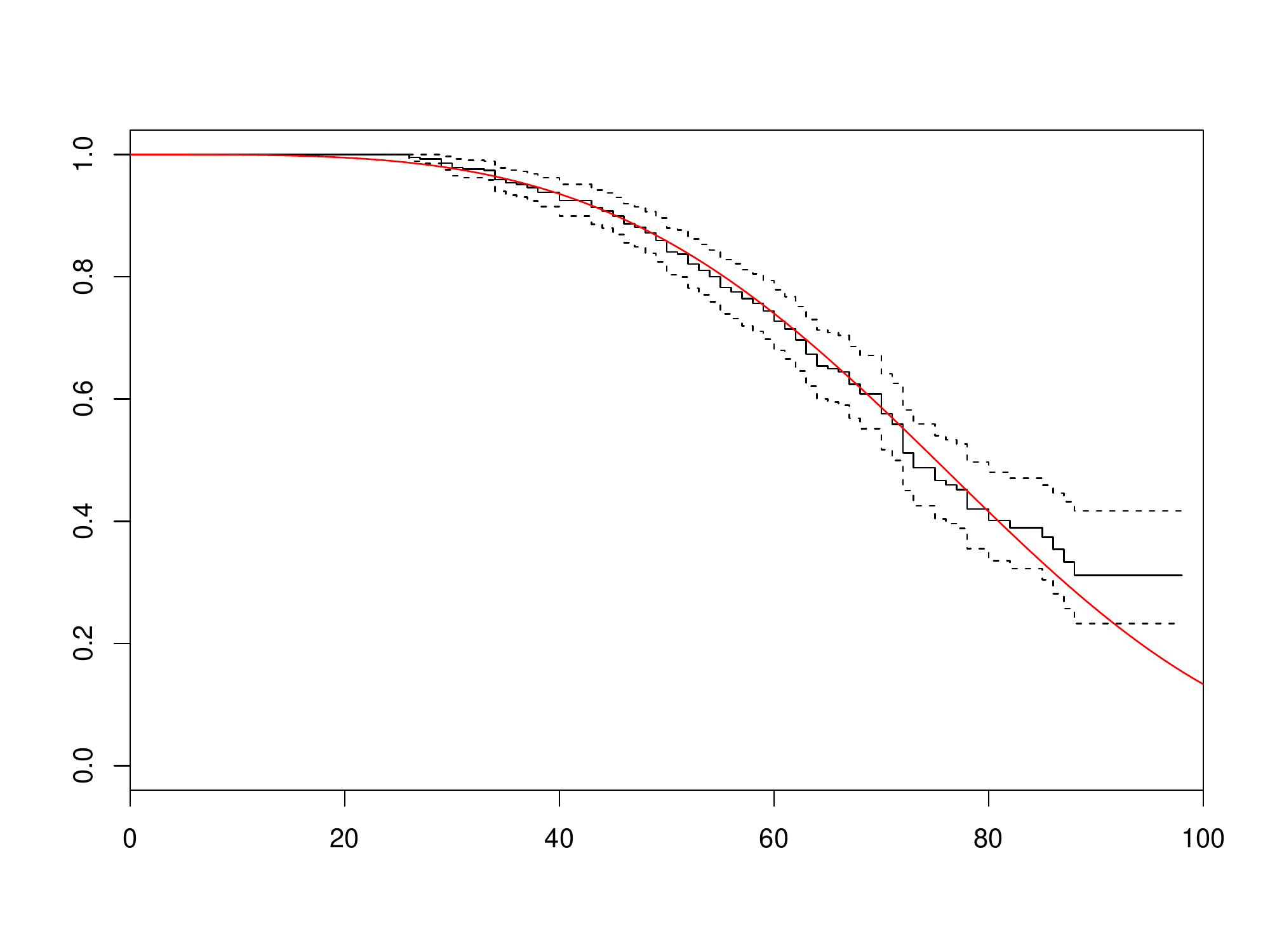}
\caption{Comparison between a Weibull parametric method to estime the Survival curve in the Swedish dataset}\label{swedish}
\end{figure}

In Swedish data, the proportionnal effect on sex was not significant with a p-value estimated to $0.43$. Figure \ref{swedish} shows estimation of the survival curve with the Kaplan-Meier estimator (black curve) and with a Weibull parametric estimation (red curve). In this case, the Weibull estimation fit almost perfectly the non-parametric curve, with the noticeable exception of the age 90 and more where the Weibull distribution clearly underestimate the survival curve. Moreover, the survival estimated in the Swedish families is higher than in Portuguese and Val30Met French families. This results are consistent with those found in \cite{hellman2008heterogeneity}

%%%%%%%%%%%%%%%%%%%%%%%%%%%%%%%%%%%%%%%%%%%%%%%%%%%%%%%%%%%%%%%%%
%                                              Discussion                                                                                 					                    %%%%%%%%%%%%%%%%%%%%%%%%%%%%%%%%%%%%%%%%%%%%%%%%%%%%%%%%%%%%%%%%%

\newpage
\section{Discussion}

In this paper, we have proposed a semi-parametric method for estimating survival functions using pedigrees with incomplete genotype information. Latent genotypes are handled by believed propagation for pedigrees and a EM algorithm allows to estimate Survival curves with weights representing the probability to carry the mutation. The method can accommodate covariates in a proportional hazards model and account for potential stratification on covariates. The believed propagation method is implemented in C++ and EM algorithm is implemented in R.

As the pedigree are ascertained through an affected individual, the proband's phenotype exclusion method is used to avoid ascertainment biais. The problem of ascertainment in segregation analysis arises when families are selected for study through
ascertainment of affected individuals. An important part of the problem is how to handle the pedigree structure, and so to model correctly ascertainment in the likelihood. Statistically, the sampling scheme can be trough as a multistage sampling method (1- one or several probands are ascertained; 2- a sequential sampling scheme is applied). Vieland and al have shown \cite{vieland1995inherent} that ``\emph{modeling the ascertainment scheme is an intractable problem}''. But she has used only sibships. This problem of ascertainment deserves more works and developments. For example, to generalize the Vieland’s approaches to arbitrary pedigrees larger than sibships and to more general problems as penetrance function estimation for diseases with variable incidence with age.

In the results Part, we have compared our non-parametric estimation to a Weibull parametric one and have seen that a Weibull parametric estimation fails to fit the survival curve estimated with our method. Additional parameters could be introduced into the Weibull model in order to improve its capacity of adjustment to the data but might involve overparametrization. Moreover, we have not been able to compare our non-parametric method to that introduced in\cite{alarcon2009ideal} based on empirical likelihood because this last method does not handle unknown genotypes.

An interesting extension of this work would be to account for the possible correlation between member of the same family by  including a frailty in the survival function. The familial frailty would typically represent an unknown shared exposure to some environmental factor or to some kind of polygenic effect. However, the estimation of such models is known to be challenging, especially in the context of non-parametric survival estimation \cite{therneau2015mixed, rondeau2012frailtypack}. Further investigation will be conducted on this important subject in our forthcoming work.

As illustration, we have estimated Survival function in three samples of different origin : French, Portuguese and Swedich families. We have notices that Survival curves had different estimation according to the origin. Moreover, in comparing our non-parametric estimation with a Weibull parametric estimation in Portuguese families, we have observed that the Weibull model did not fit well the Survival Curve. In \cite{alarcon2009pel}, Survival function was estimated with an extended Weibull model in which a parameter $\kappa$ was introduced in order to take into account the possibility that some carriers will never develop the disease and the $\kappa$ was estimated to $0.09$ with a $p < 0.001$ showing that almost 10\% of carrier will never develop the disease. We were not able to replicate this observation in the current analysis which clearly questions its relevance.

%%%%%%%%%%%%%%%%%%%%%%%%%%%%%%%%%%%%%%%%%%%%%%%%%%%%%%%%%%%%%%%%%
%%%%%%                                                                                         %%%%%%%%%%%%%%%%%%%%%%%%%%%%%%%%%%%%%%%%%%%%%%%%%%%%%%%%%%%%%%%%%%%%%%%%%%%%%%%%%%%%%%
%\vfill \eject
%\newpage
%\bibliographystyle{wileyj}
\bibliographystyle{plainnat}
\bibliography{biblio-nah.bib}

\end{document}